\documentclass{article}

\usepackage{microtype}
\usepackage{graphicx}
\usepackage{subfigure}
\usepackage{booktabs} 

\usepackage{hyperref}

\usepackage[accepted]{icml2025}

\usepackage{amsmath}
\usepackage{amssymb}
\usepackage{mathtools}
\usepackage{amsthm}
\usepackage{tcolorbox}
\usepackage[capitalize,noabbrev]{cleveref}

\theoremstyle{plain}

\theoremstyle{definition}

\theoremstyle{remark}

\usepackage[textsize=tiny]{todonotes}

\def\BibTeX{{\rm B\kern-.05em{\sc i\kern-.025em b}\kern-.08em
    T\kern-.1667em\lower.7ex\hbox{E}\kern-.125emX}}
    
        \newcommand{\red}[1]{\textcolor{red}{{#1}}}

\newcommand{\blue}[1]{\textcolor{blue}{#1}}

\begin{document}

\twocolumn[
\icmltitle{Prompting Wireless Networks: Reinforced In-Context Learning for Power Control}

\icmlsetsymbol{equal}{*}

\begin{icmlauthorlist}
\icmlauthor{Hao Zhou}{equal,yyy}
\icmlauthor{Chengming Hu}{equal,yyy}
\icmlauthor{Dun Yuan}{yyy}
\icmlauthor{Ye Yuan}{yyy}
\icmlauthor{Di Wu}{yyz}
\icmlauthor{Xue Liu}{yyy}
\icmlauthor{Jianzhong (Charlie) Zhang}{comp}
\end{icmlauthorlist}

\icmlaffiliation{yyy}{School of Computer Science, McGill University, Montreal, QC H3A 0E9, Canada.}
\icmlaffiliation{yyz}{School of Electrical and Computer Engineering, McGill University, Montreal, QC H3A 0E9, Canada.}
\icmlaffiliation{comp}{Samsung Research America, Plano, Texas, TX 75023, USA.}

\icmlcorrespondingauthor{Hao Zhou}{haozhou029@gmail.com}
\vskip 0.3in
]
\printAffiliationsAndNotice{\icmlEqualContribution}

\thispagestyle{fancy}            
\chead{\textbf{Accepted by 2025 International Conference on Machine Learning (ICML) Workshop on ML4Wireless}} 

\renewcommand{\headrulewidth}{0pt}      
\pagestyle{plain}

\begin{abstract}

To manage and optimize constantly evolving wireless networks, existing machine learning (ML)-based studies operate as black-box models, leading to increased computational costs during training and a lack of transparency in decision-making, which limits their practical applicability in wireless networks.
Motivated by recent advancements in large language model (LLM)-enabled wireless networks, this paper proposes \textbf{ProWin}, a novel framework that leverages reinforced in-context learning to design task-specific demonstration \textbf{Pro}mpts for \textbf{Wi}reless \textbf{N}etwork optimization, relying on the inference capabilities of LLMs without the need for dedicated model training or fine-tuning. 
The task-specific prompts are designed to incorporate natural language descriptions of the task description and formulation, enhancing interpretability and eliminating the need for specialized expertise in network optimization. 
We further propose a reinforced in-context learning scheme that incorporates a set of advisable examples into task-specific prompts, wherein informative examples capturing historical environment states and decisions are adaptively selected to guide current decision-making.   
Evaluations on a case study of base station power control showcases that the proposed ProWin outperforms reinforcement learning (RL)-based methods, highlighting the potential for next-generation future wireless network optimization.

\end{abstract}

\section{Introduction}

The envisioned 6G network is expected to become increasingly complex, encompassing diverse application scenarios and advanced signal processing techniques such as mmWave and THz networks, reconfigurable intelligent surfaces, near-field communications, and movable antennas, among others~\cite{zhang20196g}.   
To optimize such constantly evolving networks, convex optimization-based methods commonly formulate a dedicated problem for each specific task and transform the objective functions or constraints into a convex form~\cite{10636212}. 
On the other hand, although machine learning (ML)-based methods, such as reinforcement learning~\cite{zhou2023survey, burkart2021survey}, require less stringent problem formulations and exhibit impressive performance across diverse tasks, the tedious model training and fine-tuning demand extensive computational resources, making them impractical for wireless networks, particularly in resource-constrained environments. 
Additionally, existing ML-based methods operate as black boxes, lacking interpretability and transparency in the decision-making process, which poses challenges in wireless network-related applications.
Hence, the above challenges, including redundant problem formulation, computationally intensive demands, and lack of transparency, are calling for more advanced methods to optimize and manage next-generation wireless networks.

Generative AI (GenAI), particularly large language models (LLMs), has recently attracted considerable attention in the field of natural language processing, offering promising opportunities to develop LLM-enabled wireless networks across diverse tasks, including edge intelligence~\cite{lin2023pushing}, semantic communication~\cite{park2024towards, chen2024personalizing}, and network design~\cite{qiu2024large}, among others. 
Specifically, in-context learning leverages the inference capabilities of LLMs by incorporating natural language-based demonstration prompts to guide task execution~\cite{dong2022survey}. 
By following task-specific descriptions within demonstration prompts, LLMs can identify the task type using knowledge embedded in pre-trained corpora and adopt effective task-solving strategies~\cite{wies2023learnability, 10356715, ouyang2022training}.  
Given the high computational efficiency without dedicated model training or fine-tuning, in-context learning presents a promising method in wireless networks, particularly in resource-constrained environments, by reducing the energy consumption and alleviating computational burden on network resources.
Moreover, in-context learning enables natural language-based task design and implementation through demonstration prompts, allowing operators to easily formulate problems using human language and instructions~\cite{min2022rethinking}. 
This user-friendly method holds great potential to reduce human effort and lower the need for specialized expertise in wireless networks.  
In-context learning also offers a distinct advantage in explainable decision-making process by generating natural language-based justifications for outputs~\cite{bariah2023understanding}, thereby enhancing the transparency and interpretability of the reasoning process and enabling operators to clearly understand and manage wireless networks in the 6G era~\cite{maatouk2024large}.

To this end, this paper proposes \textbf{ProWin}, a framework that leverages reinforced in-context learning to design task-specific demonstration \textbf{Pro}mpts for \textbf{Wi}reless \textbf{N}etwork optimization, eliminating the need for model training or fine-tuning.
Compared to existing LLM-enabled wireless network studies~\cite{su2024federated, yan2025federated, park2024towards, qiu2024large}, our proposed ProWin addresses a more complex and dynamic network optimization problem, enabling the LLM to make informed decisions by learning from and adapting to varying network observations. 
The task-specific prompts are first designed to incorporate natural language descriptions of task description related to task objectives, definitions, and rules, making it efficient to formulate tasks while reducing the need for specialized knowledge in network optimization.
To better optimize dynamically evolving network environments, we further propose a reinforced in-context learning scheme that adaptively embeds a set of advisable examples into task-specific prompts. 
Specifically, each example comprises a historical environment state, the decision made in that state, and the corresponding reward evaluated after the decision.
Moreover, to effectively select relevant examples for the current task, we introduce state-based and ranking-based schemes for discrete and continuous states, respectively, ensuring that the most informative past examples are selected to provide valuable guidance for current decision-making.
In this way, the LLM is expected to make decisions based on task-specific prompts embedded with relevant examples. The resulting decision, current environment state, and corresponding evaluated reward are then stored as a new example in an experience pool, serving as a reference for future decision-making.
In this paper, we consider base station (BS) power control as a case study, which is a fundamental and critical optimization task in wireless networks that has been extensively studied with diverse algorithms, including convex optimization, game theory, and reinforcement learning, among others. 
To summarize, the \textbf{main contributions} of our work are provided as follows:
\begin{itemize}
    \item We propose ProWin, a novel LLM-enabled method that designs task-specific demonstration prompts for wireless network optimization, eliminating the need for model training or fine-tuning, and enabling  efficient task formulation with reduced reliance on specialized expertise in network optimization.  
    \item We introduce a reinforced in-context learning scheme that adaptively selects and embeds a set of advisable examples into task-specific prompts, ensuring effective network optimization in dynamically evolving environments through guidance from the selected examples. 
    \item We conduct comprehensive experiments on the case study of base station power control, demonstrating  the consistent superiority of the proposed across ProWin various LLMs.  
\end{itemize}

\section{System Model}

\subsection{Problem Formulation}
This section introduces a BS power control problem, serving as a case study to demonstrate the application of the proposed ProWin for wireless network optimization. 
Considering a BS with $U_b$ users, the achievable data rate $C_{b,u}$ between BS $b$ and user $u$ is defined by \cite{zhou2022learning}
\begin{equation}
\label{eq4}
\resizebox{0.91\hsize}{!}{$
C_{b,u}=\sum\limits_{k=1}\limits^{K_{b}}d_{k}log(1+ \frac{p_{b,k}h_{b,k,u}\gamma_{b,k,u}}{\sum\limits_{b'\in B_{-b}}{p_{b',k'}h_{b',k',u'}\gamma_{b',k',u'}}+d_{k}N_{0}}),$}
\end{equation}
where $K_{b}$ is the total number of resource blocks (RBs) in BS $b$, $d_{k}$ is the bandwidth of RB $k$, 
$p_{b,k}$ indicates the transmission power of BS $b$ on RB $k$, 
$h_{b,k,u}$ defines the channel gain between BS $b$ and user $u$ on RB $k$, and $N_{0}$ is the noise power density. 
For the RB allocation, $\gamma_{b,k,u} \in \{0,1\}$ indicates whether RB $k$ is allocated to the transmission for user $u$.  
For the interference, $B_{-b}$ represent the set of adjacent BSs except for BS $b$, $p_{b',k'}h_{b',k',u'}\gamma_{b',k',u'}$ defines the inter-cell interference, and we assume orthogonal frequency-division multiplexing is applied to eliminate intra-cell interference.

This work aims to minimize the BS transmission power and meanwhile satisfy the average data rate constraint \cite{chiang2008power}:
\begin{subequations}\label{e5:main}
\begin{align}
\min_{P_{b}} & \enspace \sum\nolimits_{b\in B} P_{b}     & \tag{\ref{e5:main}} \\
\text{s.t.}  & \enspace 0 \leq P_{b} \leq P_{max},  & \label{e5:a}\\
& \enspace  P_{b}=\sum\nolimits_{k=1}^{K_{b}} p_{b,k},  & \label{e5:b}\\ 
& \enspace \sum\nolimits_{u=1}^{U_{b}}C_{b,u}/U_{b}  \geq C_{min}, & \label{e5:c}   
\end{align}
\end{subequations}
where $P_b$ is the total transmission power of BS $b$ and $P_{b}=\sum\nolimits_{k=1}^{K_{b}} p_{b,k}$, $p_{b,k}$ has been defined in equation (\ref{eq4}) as the transmission power of RB $k$, 
$P_{max}$ is the maximum power, 
$U_b$ is the total number of users, and $C_{min}$ is the average achievable data rate constraint.
We assume $P_b$ is equally allocated to all RBs, and a proportional fairness method is used for RB allocation, which has been widely used as a classic approach. Then we can better focus on LLM features. 

%

\subsection{Language-based Power Control Task Description}
\label{sec-task}

Problem (\ref{e5:main}) has been extensively investigated in existing studies, but this work differs from previous works by presenting a unique view from the perspective of LLM-enabled network optimization. 
Instead of defining specific equations as in (\ref{e5:main}), here we use natural language to describe the optimization task: 1) Firstly, it inherently avoids the complexity of defining dedicated problem formulations, which is usually time-consuming;  
2) Secondly, language-based task description is a user-friendly approach, and network operators can easily formulate the task without requiring any professional optimization-related knowledge.

In this power control case study, the task description involves “$Task\_goal$”, “$Task\_definition$”, and extra “$Rules$”. The defined task description is shown below, which will further be used to prompt LLMs:
\begin{tcolorbox}[title = {Task description for BS transmission power control}]
\label{box1}
\textbf{Task goal}: You have a decision-making task for base station power control, and you need to select between 4 power levels from 1 to 4.\\
\textbf{Task definition}: You have to consider the specific user number of each case, which is the “base station user number”.\\
Following are some examples $\{Example\_set\}$.\\
Now I will give you a new condition to solve, the current BS user number is $\{Num\_BS\_user\}$. \\
\textbf{Rules}: Now please select from “level 1”, “level 2”, “level 3”, and “level 4” based on the above examples.
\end{tcolorbox}

In particular, the $Task\_goal$ first specifies a “\textit{decision-making task for base station power control}”, and the goal is to “\textit{select between 4 power levels}”\footnote{Here we select 4 power levels as an example, which can be changed to any number of levels}. Then the $Task\_definition$ introduces the environment states we need to consider. For example, this work assumes the total user numbers may change dynamically, and then the LLM has to consider the “\textit{user number}” of each case. 
%
After that, the example set $\mathcal{E}_{t}$ is included by “\textit{Following are some examples....}”, and we provide a new condition for the LLM to solve with the current user number $U_b$. 
Finally, we set extra reply rules such as “\textit{select from ... based on the above examples}”, indicating the LLM to focus on decision-making.

\begin{figure*}[t]
\centering
\includegraphics[width=1\linewidth]{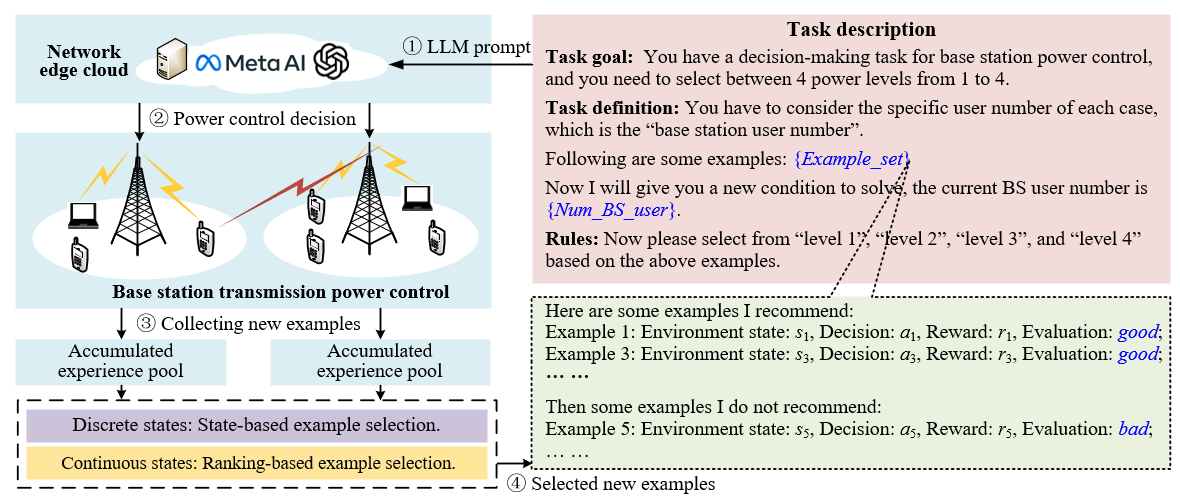}
\caption{Overall design of the proposed LLM-enabled in-context learning for transmission power control.}
\label{fig-optimization}
\end{figure*}

\section{In-context Learning-Based Optimization Algorithm}
\label{sec-conext}

This section will introduce the proposed in-context learning algorithm, aiming to optimize power control by using natural language-based task descriptions in Section \ref{sec-task}.

\subsection{In-context Learning}
\label{sec-incontext}
In-context learning refers to the process that LLMs can learn from formatted natural language such as task descriptions and task solution demonstrations, to improve the performance on target tasks.
In-context learning can be defined as \cite{dong2022survey}
\begin{equation}
\label{eq_llm}
D_{task} \times \mathcal{E}_{t} \times s_{t} \times \mathcal{LLM} \Rightarrow a_{t},
\end{equation}
where $D_{task}$ is the task description and query, $\mathcal{E}_{t}$ is the set of examples at time $t$, $s_{t}$ is the environment state at time $t$ that is associated with the target task,
$\mathcal{LLM}$ indicates the LLM model, and $a_{t}$ is the LLM output. 
Here we expect the LLM can utilize the initial task description $D_{task}$, learn from the example set  $\mathcal{E}_{t}$, and then make decision $a_{t}$ based on current environment state $s_{t}$ of the target task.

The LLM's in-context learning capabilities can be considered as implicit fine-tuning according to \cite{dai2022can}. 
%
LLMs will produce meta-gradients based on given examples $\mathcal{E}$ by forward computation, and then the meta-gradients are applied by using the attention mechanism to build an in-context learning model.
Specifically, consider $x$ as the input representation of a query token $t$, and $\textbf{q} = W_{Q}x$ as the query vector in the attention mechanism. Then the attention of a head is formulated as
\begin{equation}\label{eq-atten}
\begin{aligned}
f_{ICL}(\textbf{q})= & {\rm attention}(\textbf{q},K,V) \\
= & {\rm softmax}(\frac{\textbf{q}(W_{K}[\mathcal{Q};\mathcal{E}])^{T}}{\sqrt{d}})W_{V}[\mathcal{Q};\mathcal{E}]
\end{aligned}
\end{equation}
where $W_Q$, $W_K$ and $W_V$ represent the weight matrices of attention queries, keys, and values, respectively; $d$ is the scaling factor; $\mathcal{Q}$ and $\mathcal{E}$ denote the input representations of query tokens and demonstration tokens, respectively, and $[\mathcal{Q};\mathcal{E}]$ is the concatenated matrix of $\mathcal{Q}$ and $\mathcal{E}$. 
For ease of qualitative analysis, we relax the standard attention to linear attention by removing the softmax operation and the scaling factor from equation (\ref{eq-atten}):
\begin{equation}
\begin{aligned} \label{eq-zcl}
f_{ICL}(\textbf{q}) \approx & \textbf{q} (W_{K}[\mathcal{Q};\mathcal{E}])^{T} W_{V}[\mathcal{Q};\mathcal{E}]  \\
= & \textbf{q} (W_{K}\mathcal{Q})^{T}  W_{V} \mathcal{Q}  +  \textbf{q} (W_{K}\mathcal{E})^{T}  W_{V} \mathcal{E} \\
= & \tilde{f}_{ICL}(\textbf{q})
\end{aligned}
\end{equation}

We define $W_{ZSL}=(W_{K}\mathcal{Q})^{T} W_{V}\mathcal{Q}$ for ease of notations, since the $W_{ZSL}$ parameters include the query token $\mathcal{Q}$ only. 
Therefore, $W_{ZSL}$ indicates the zero-shot learning case without examples. Then we can rewrite equation (\ref{eq-zcl}) as 
\begin{equation}
\begin{aligned} \label{eq-linear}
    \tilde{f}_{\text{ICL}}&(\mathbf{q})= \mathbf{q} W_{\text{ZSL}}  + \mathbf{q} \left( W_K \mathcal{E} \right)^T W_V \mathcal{E}  \\
    &= \mathbf{q} W_{\text{ZSL}}  + \text{LinearAttention} \left( \mathbf{q}, W_K \mathcal{E}, W_V \mathcal{E} \right) \\
    &= \mathbf{q} W_{\text{ZSL}}  + \mathbf{q} \sum_i  \left( \left( W_K E_i \right)^T W_V E_i  \right) \\
    &= \mathbf{q} W_{\text{ZSL}}  +  \mathbf{q} \sum_i \left( \left( W_K E_i \right) \otimes \left( W_V E_i \right) \right) \\
    &= \mathbf{q} W_{\text{ZSL}} + \mathbf{q} \Delta W_{\text{ICL}}\\
    &= \mathbf{q} \left( W_{\text{ZSL}} + \Delta W_{\text{ICL}} \right) .
\end{aligned}
\end{equation}
Here $\Delta W_{\text{ICL}}=\sum_i \left( \left( W_K E'_i \right) \otimes \left( W_V E'_i \right) \right)$, which indicates the updated weight when examples $E \in \mathcal{E}$ are provided. 
$\mathbf{q} \left( W_{\text{ZSL}} + \Delta W_{\text{ICL}} \right)$ also aligns with the weight updating of back-propagation algorithm. It proves that the examples $\mathcal{E}$ will affect the in-context learning weight $\Delta W_{\text{ICL}}$, and therefore LLMs can learn from examples and generate replies.

\subsection{Examples and Optimization Framework Design}
The analyses in Section \ref{sec-incontext} show that 
examples are of great importance in in-context learning, which will directly affect the $\Delta W_{\text{ICL}}$ values. 
However, many network optimization problems have continuous environment states, e.g., adjusting the BS transmission power based on user-BS distance. Such cases mean that there may be an infinite number of examples, and therefore identifying the most relevant and useful examples becomes challenging. Here we define an example by
\begin{equation}
\label{eq_example}
E=\{s, a, r(s,a)\}, E \in \mathcal{E},    
\end{equation}
where $s$ and $a$ are environment state and decision, respectively. Inspired by reinforcement learning, we further define a reward value to evaluate the decision $a$ by
\begin{equation}
\label{eq_reward}
r=P_{target}-P_{b}-\beta,   
\end{equation}
where $P_{target}$ is a target power consumption, and $P_b$ has been defined in problem (\ref{e5:main}) as the total power consumption of BS $b$. $\beta$ is a penalty term, which is only applied when constraint  (\ref{e5:c}) is not satisfied. Then, $r$ provides a comprehensive metric to evaluate the selected decision $a$ under environment state $s$.

Fig.\ref{fig-optimization} shows the overall design of the proposed in-context learning algorithm for transmission power control. 
Specifically, the above task description $D_{task}$, current environment state $s_t$, and selected examples $\mathcal{E}_{t}$ are integrated as input prompt as defined in equation (\ref{eq_llm}), and then the LLM will generate a power control decision $a_t$ based on $s_t$ and the experiences in $\mathcal{E}_t$.
Then, the decision $a_t$ is implemented, the achieved data rate $C_{b,u}$ is collected, and the reward $r_t$ is calculated as equation (\ref{eq_reward}). $E_{t}=\{s_t, a_t, r_{t}(s_t,a_t)\}$ becomes a new example in the accumulated experience pool $\mathcal{E}_{pool}$ in Fig.\ref{fig-optimization}.   
After that, based on the next environment state $s_{t+1}$, a new example set $\mathcal{E}_{t+1}$ is selected, and the selected examples are inserted into the task description with $s_{t+1}$, becoming a new prompt for the LLM model to generate $a_{t+1}$.

\begin{algorithm*}[tb]
   \caption{Proposed In-context Learning-based Algorithm for Network Optimization}
   \label{alg:example}
\begin{algorithmic}
   \STATE {\bfseries Input:} Network parameters: BS and user locations, BS transmission power constraint $P_{max}$, user data rate constraint $C_{min}$. Optimization parameters: penalty term $\beta$, $\epsilon$ for epsilon-greedy policy, weighting factor $\tau$. Initialize the experience pool $\mathcal{E}_{pool}$.  
   \STATE \textbf{Designing} input prompt: Task goals, task definition, and rules as in Section \ref{sec-task}. 
   \REPEAT
    \IF{$rand(0,1) < \epsilon $}
   \STATE Selecting transmission power level randomly.
   \ELSE  
        \IF{Discrete state problem}
        \STATE Given current state $s$, selecting relevant examples $\mathcal{E}_{relevant}$ using equation (\ref{eq_example_re}) from experience pool $\mathcal{E}_{pool}$.
        \ELSIF{Continuous state problem}  
        \STATE Calculating the $\mathcal{L}(E,s_{target})$ metric of all examples in $\mathcal{E}_{pool}$ using equation (\ref{eq-metric}), and then selecting the top-K relevant examples $\mathcal{E}_{relevant}$.    
        \ENDIF
        \STATE Inserting the selected examples $\mathcal{E}_{relevant}$ into the prompting template. Feeding the prompts to LLMs $D_{task} \times \mathcal{E}_{t} \times s_{t} \times \mathcal{LLM} \Rightarrow a_{t}$, and generating the selected power level $a_{t}$.
   \ENDIF 
   \STATE Inserting the new example $E=\{s, a, r(s,a)\}$ into the accumulated experience pool $\mathcal{E}_{pool}$.
   \UNTIL{Reaching the max number of iterations or the result converges.}
\end{algorithmic}
\end{algorithm*}

\subsection{State-based Example Selection for Discrete State Problems}
\label{sec-discrete}
Selecting appropriate examples is critical for in-context learning\cite{chen2023many}.
For problems with discrete environment states, relevant demonstrations can be easily identified by finding existing examples with the same states in the accumulated experience pool $\mathcal{E}_{pool}$.  
Considering a target task with environment state value $s_{target}$, the set of relevant examples can be identified by
\begin{equation}
\label{eq_example_re}
\mathcal{E}_{relevant}= \Big \{E\{s,a,r(s,a)\} \Big| s=s_{target}, E \in \mathcal{E}_{pool} \Big\},
\end{equation}
where $\mathcal{E}_{pool}$ is the accumulated experience pool in Fig. \ref{fig-optimization}. 
Given $\mathcal{E}_{relevant}$, we can easily select recommended examples with high reward, i.e., top-K examples, and inadvisable examples, e.g., examples with lower reward or violating the minimum data rate constraint.   

In addition, we include a well-known epsilon-greedy policy to balance exploration and exploitation. 
\begin{equation} \label{eq_epsilon}
a=\left\{
\begin{array}{lcl} \text{Random action selection},     & \text{if }  rand < \epsilon; \\
\text{LLM-based decision-making,} &  \text{else},
\end{array} \right.
\end{equation}
where $\epsilon$ is a predefined value, and $rand$ is a random number between 0 and 1. 
Therefore, the random exploration in equation (\ref{eq_epsilon}) can constantly explore new examples, and then the LLM model can learn from better relevant examples $\mathcal{E}_{relevant}$ to improve the performance.

\subsection{Ranking-based Example Selection for Continuous State Problems}
Compared with discrete-state problems, environments with continuous states can be much more complicated. 
For instance, when using average user-BS distance as an environment state for BS transmission power control with a target task $s_{target}$, it is unlikely to find a specific existing example $E\{s,a,r(s,a)\}$ with $s=s_{target}$, since $s_{target}$ is a random number within the BS maximum coverage distance. This problem may be solved by discretizing the continuous states into some discrete values, but this may still lead to a large number of states or extra errors. 
To this end, we define a new metric $\mathcal{L}$ for example selection with continuous states:
\begin{equation} \label{eq-metric}
\mathcal{L}(E,s_{target})=r(s,a)-\tau||s-s_{target}||,    
\end{equation}
where $\mathcal{L}(E,s_{target})$ is a comprehensive metric to evaluate the usefulness of $E=\{s, a, r(s,a)\}$ to the decision-making of $s_{target}$, 
and $||s-s_{target}||$ is the $L^2$ norm to define the distance between $s$ and $s_{target}$. 
Equation (\ref{eq-metric}) aims to jointly consider the reward and states of example $E$, and $\tau$ is a weighting factor to balance the importance of higher reward $r(s,a)$ and more similar states between $s$ and $s_{target}$. 
Specifically, a higher reward $r(s,a)$ indicates that $E$ includes a good action selection $a$ under environment state $s$, 
and meanwhile lower $||s-s_{target}||$ value means the environment state $s$ in $E$ is more similar to $s_{target}$.
Therefore, we use $\mathcal{L}(E, s_{target})$ as a comprehensive metric, and then the recommended and inadvisable examples can be selected similarly as in Section \ref{sec-discrete} by selecting the top-K examples with highest $\mathcal{L}(E,s_{target})$ value.

Finally, the proposed algorithm can be summarized as Algorithm \ref{alg:example}. With the epsilon-greedy policy, the algorithm can try different actions by random exploration, finding better network optimization decisions. 
On the other hand, LLM can utilize the accumulated experience and examples to make good decisions, balancing exploration and exploitation.  
Such a design indicates that LLMs can constantly explore the network environment and improve their decisions iteratively. 
Compared with other ML-enabled optimization techniques, the proposed algorithm has no need to update LLM model parameters. Therefore, it is more efficient with much lower complexity.


\begin{figure}[t]
\centering
\includegraphics[width=0.95\linewidth]{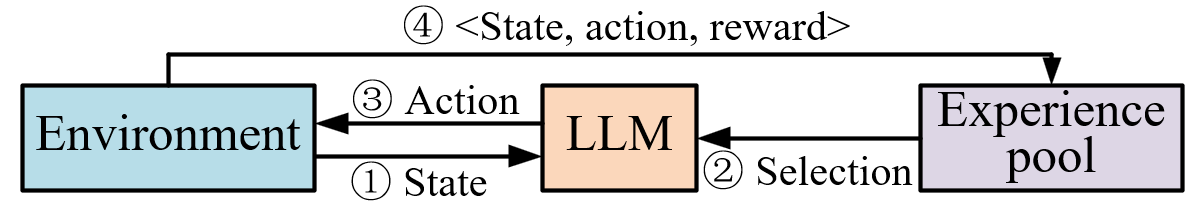}
\caption{The overall procedure of the example-related scheme.}
\label{fig-example}
\end{figure}

\subsection{Computational Complexity Analyses}

Fig. \ref{fig-example} summarizes the overall procedure of example-related schemes. In particular, the LLM will receive the state from the environment, and then use the examples provided by the experience pool to select actions such as the transmission power level. The implementation results will become a new example for the pool. 
Meanwhile, no additional computational cost is incurred for example selection, as each new example is simply appended to the accumulated experience pool after implementation. 
Secondly, for example selection in discrete state problems, it is easy to search the experience pool to identify $s=s_{target}$. For continuous states, we calculate the $\mathcal{L}(E,s_{target})$ metric for all examples in the pool, and then select the best examples accordingly. 
Therefore, the cost of example selection follows a linear complexity.
Finally, note that the LLM inference time is affected by model architecture, hardware constraints, and task types, and it can also be further optimized by quantization, sparsity exploitation, and architectural innovations.

\section{Performance Evaluation}

\subsection{Simulation Settings}
We consider three adjacent small base stations (SBSs); the user number of each SBS randomly changes from 5 to 15, and the SBS's coverage is 20 meters. The channel gain applies 3GPP urban network models, and
2 cases are evaluated:
\textbf{Case I}: Discrete states defined by user numbers of each SBS;
\textbf{Case II}: Continuous states defined by average user-SBS distance, which represents 2 kinds of network optimization problems.

Then, the simulation considers 2 main approaches:\\ 
\textbf{1) LLM-based method} includes 3 models: Llama3-8b-instruct, Llama3-70b-instruct, and GPT-3.5 turbo.
Llama3-8b is a small-scale LLM, while Llama3-70b and GPT-3.5 turbo are large models.
Using LLM models with various sizes can better evaluate the capabilities of our proposed algorithms\cite{oh2024uncertainty}.
The first 40\% episodes in the simulation are exploration phases, while the rest of the episodes are the exploitation phase.\\
\textbf{2) DRL-based method}: with dedicated model training, here we consider DRL as an optimal baseline since its capability has been demonstrated in many existing studies \cite{zhang2020deep,zhou2022learning}. 
The Markov decision process (MDP) for deep Q-learning is: states can be easily defined by considering discrete/continuous states as introduced above, actions indicate the BS transmission power levels, and rewards are defined as equation (\ref{eq_reward}).

\begin{figure}[t]
\centering
\includegraphics[width=8cm,height=6cm]{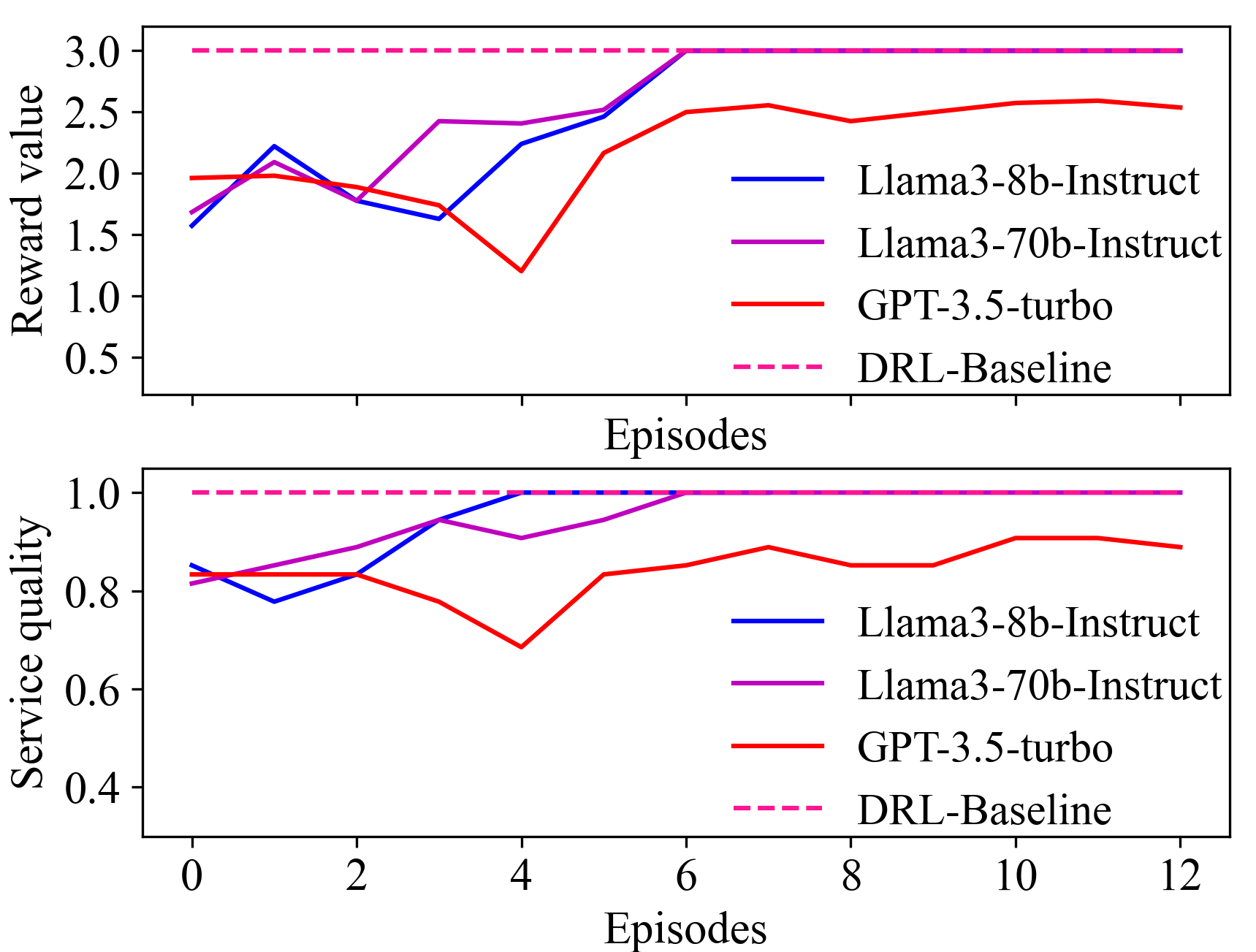} 
\caption{Discrete state space: System reward and service quality comparison of various LLMs.}
\label{f-r1}
\end{figure}

\begin{figure}[t]
\centering
\includegraphics[width=8cm,height=6cm]{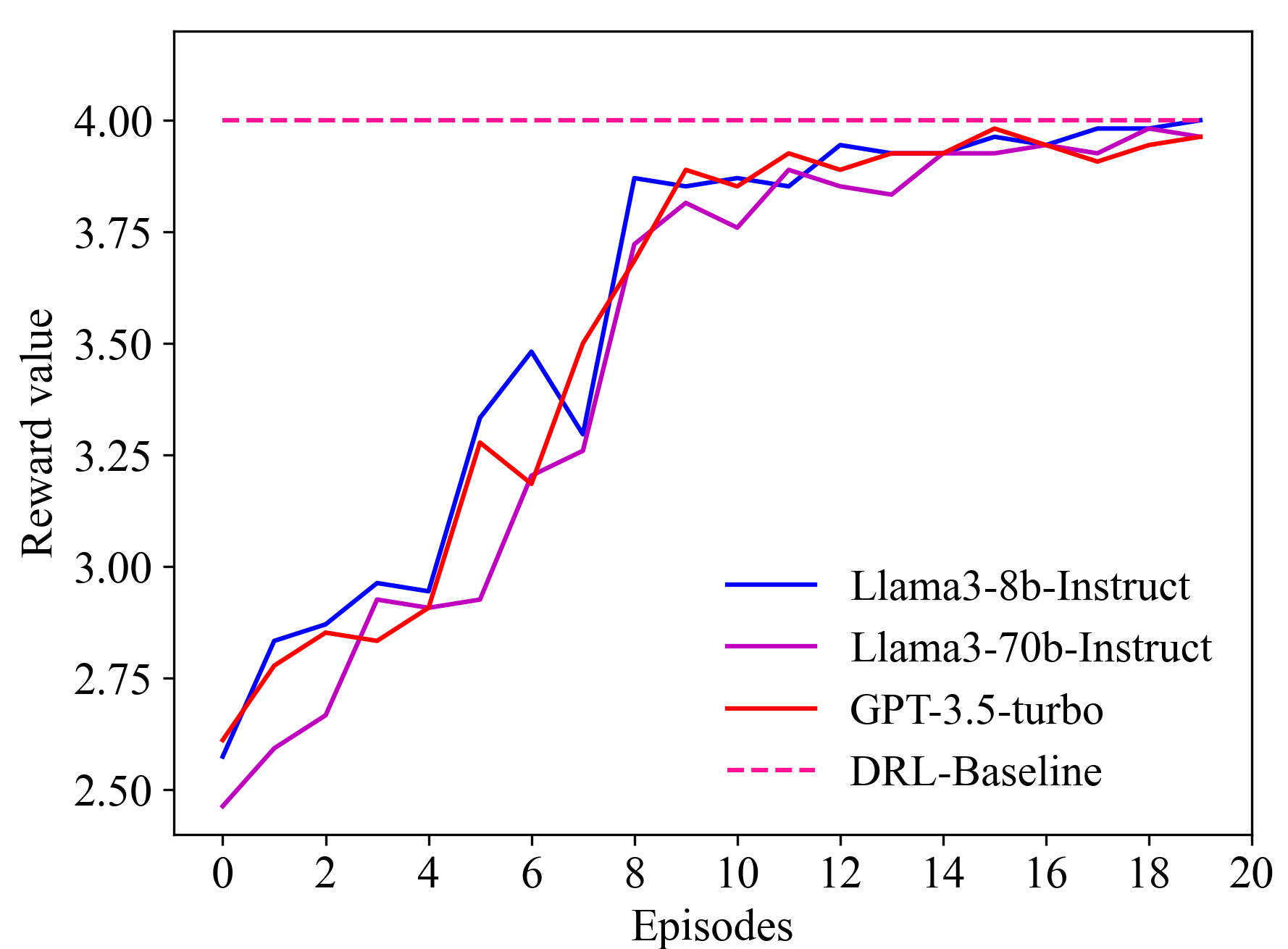} 
\caption{Continuous state space: System reward comparison of various LLMs.}
\label{f-r2}
\end{figure}

\subsection{Simulation Results}
\begin{figure}[t]
\centering
\includegraphics[width=8cm,height=6cm]{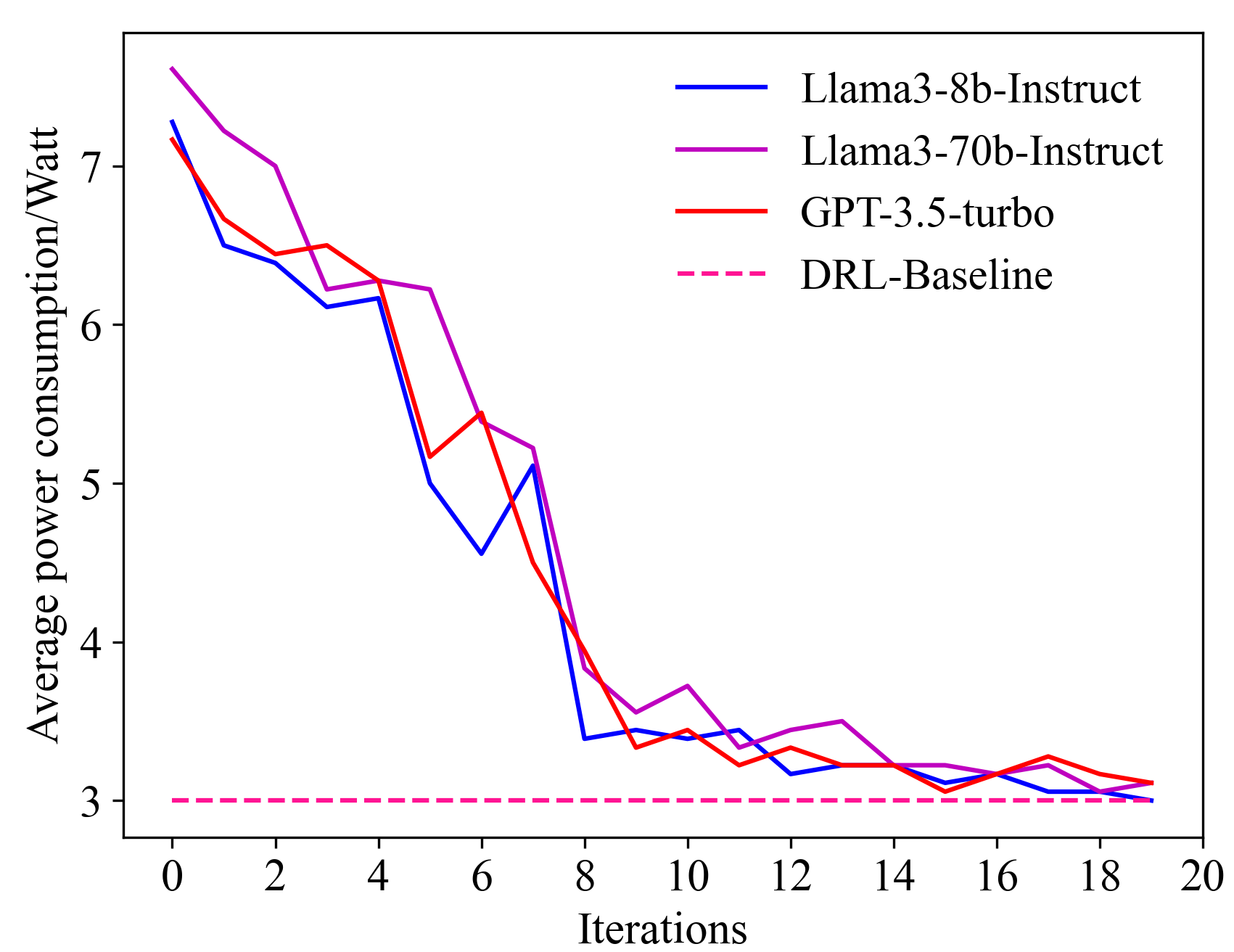} 
\caption{Continuous state space: Power consumption comparison of various LLMs.}
\label{f-r3}
\end{figure}

\begin{figure}[t]
\centering
\includegraphics[width=8cm,height=6cm]{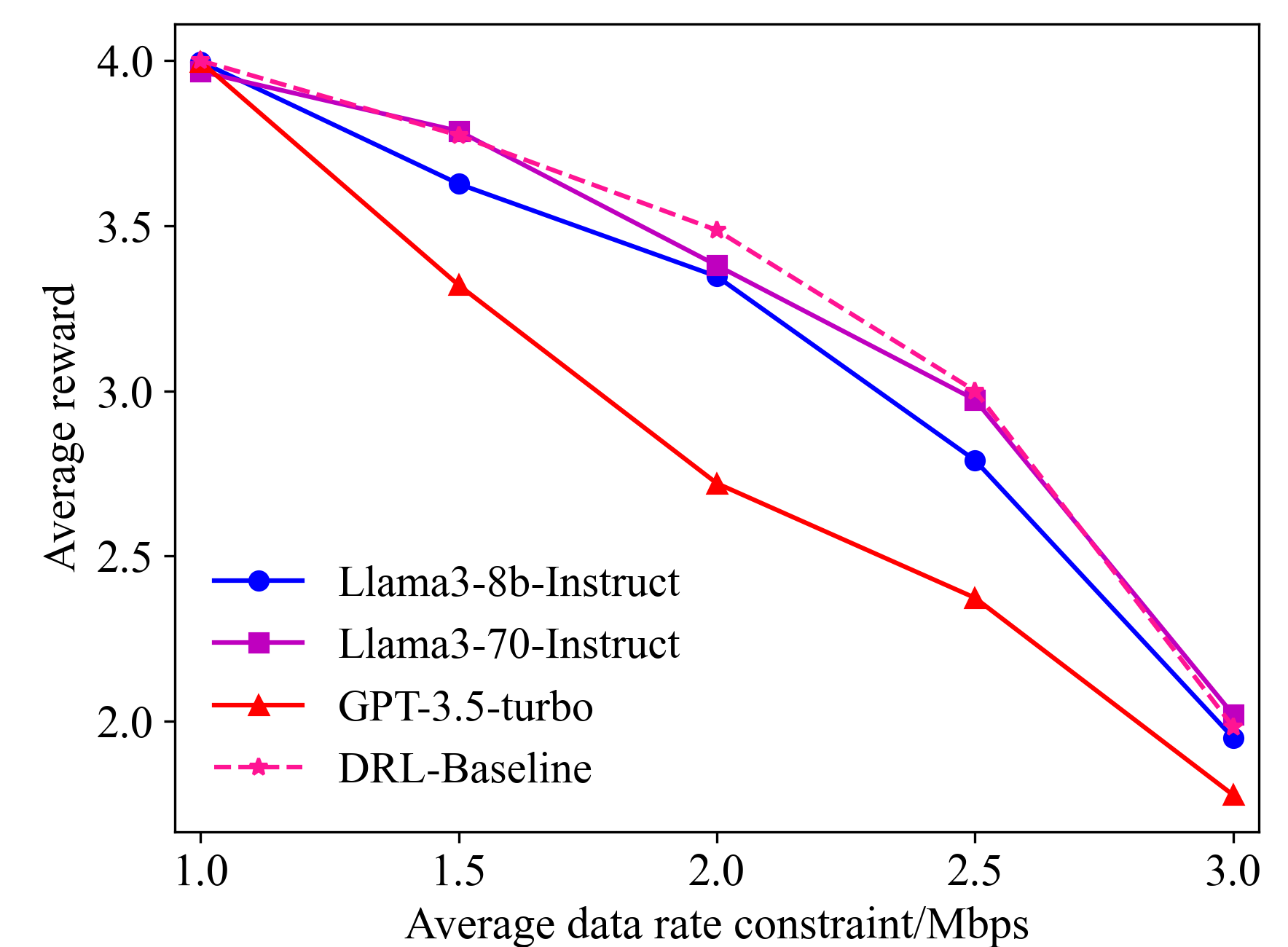} 
\caption{Continuous state space: Average reward comparison under different data rate constraints.}
\label{f-r4}
\end{figure}

\begin{figure}[t]
\centering
\includegraphics[width=8cm,height=6cm]{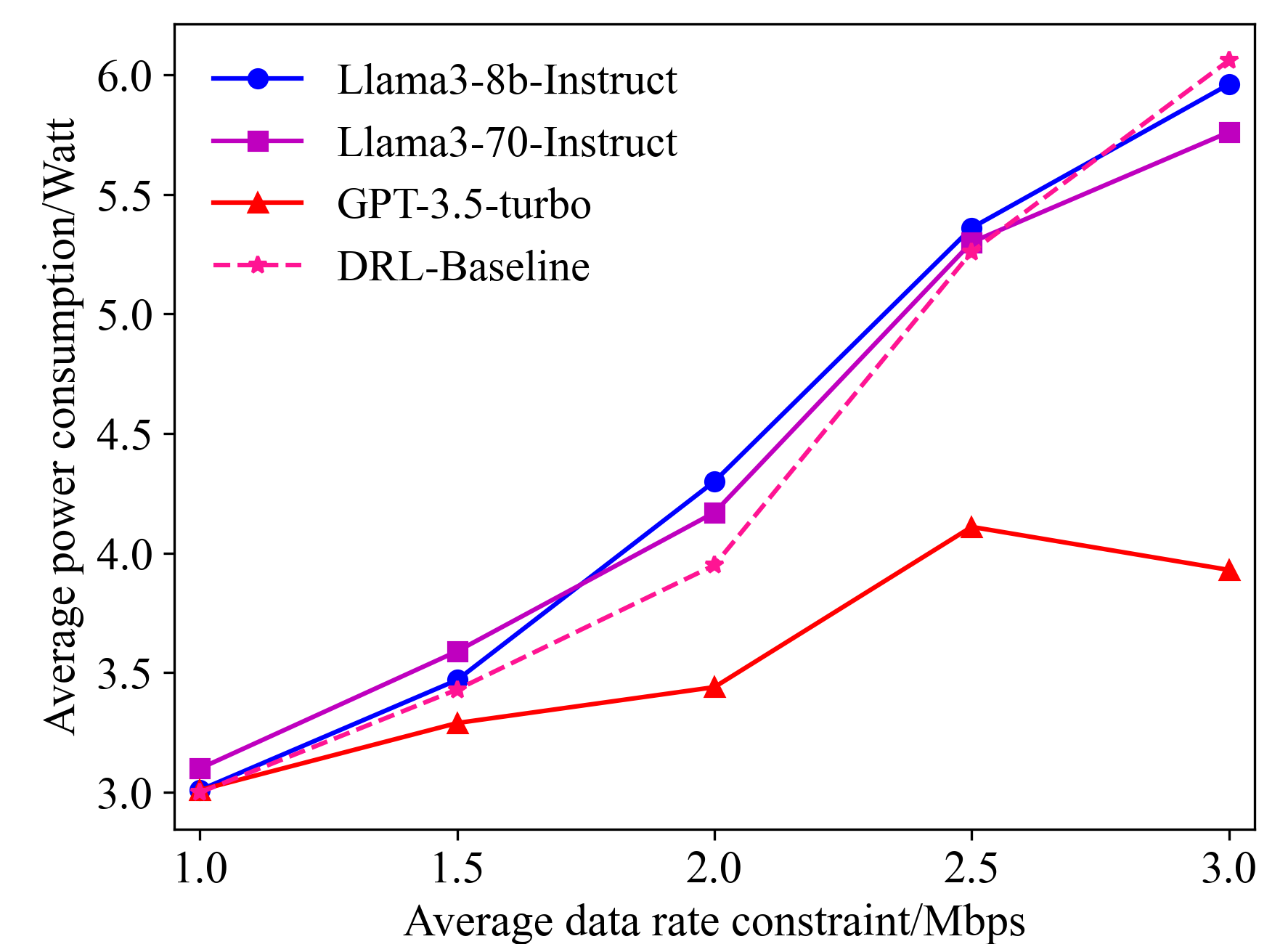} 
\caption{Continuous state space: Average power consumption comparison under different data rate constraints.}
\label{f-r5}
\end{figure}

\begin{figure}[t]
\centering
\includegraphics[width=8cm,height=6cm]{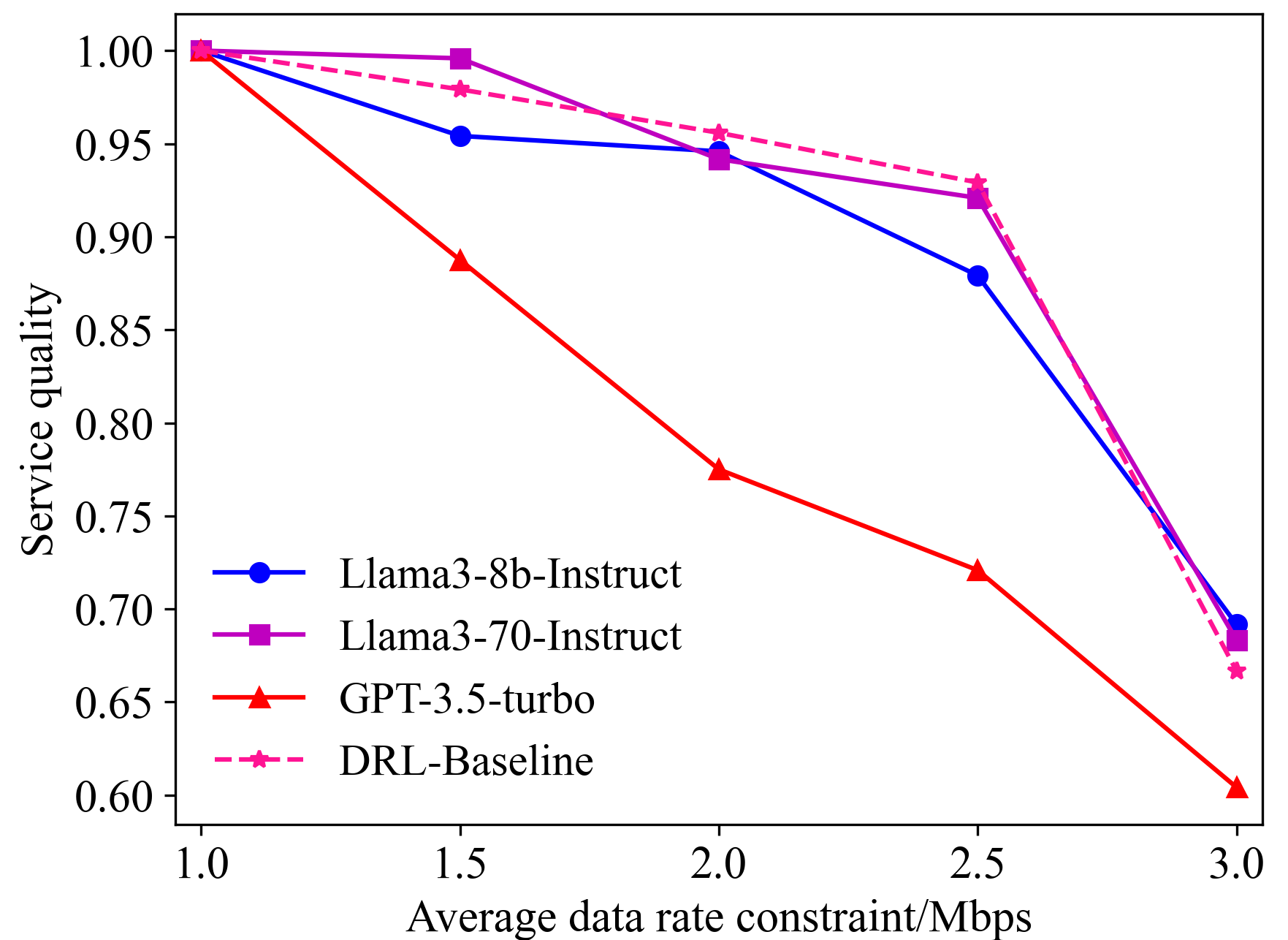} 
\caption{Continuous state space: Average service quality comparison under different data rate constraints.}
\label{f-r6}
\end{figure}

Fig. \ref{f-r1} to \ref{f-r6} show the simulation results and comparisons.

\textbf{1) Discrete State Problems}: Firstly, for discrete state problems, Fig. \ref{f-r1} presents the system reward and service quality of different LLMs. 
One can observe that both Llama3 LLMs achieve a comparable reward and service quality as the DRL baseline, while GPT-3.5 shows a lower reward and service quality.
Fig. \ref{f-r1} demonstrates that the proposed in-context learning algorithm can provide satisfactory performance for problems with a limited number of environment states.

\textbf{2) Continuous State Problems}:  Then, we consider more complicated scenarios with continuous states defined by the average user-BS distance. Fig. \ref{f-r2} and \ref{f-r3} show that all LLM models achieve higher rewards and lower power consumption as the number of episodes increases and finally converge to stable values, e.g., reward value 4 and 3 Watts average power consumption.
Specifically, with the epsilon-greedy policy, the LLM can randomly explore different actions and meanwhile make optimal decisions based on existing knowledge. 
The results demonstrate that LLMs can learn from previously accumulated examples  and then improve their performance on target tasks.
Such an iterative optimization approach is a crucial skill in addressing many real-world problems.

\begin{figure}[t]
\centering
\includegraphics[width=8cm,height=6cm]{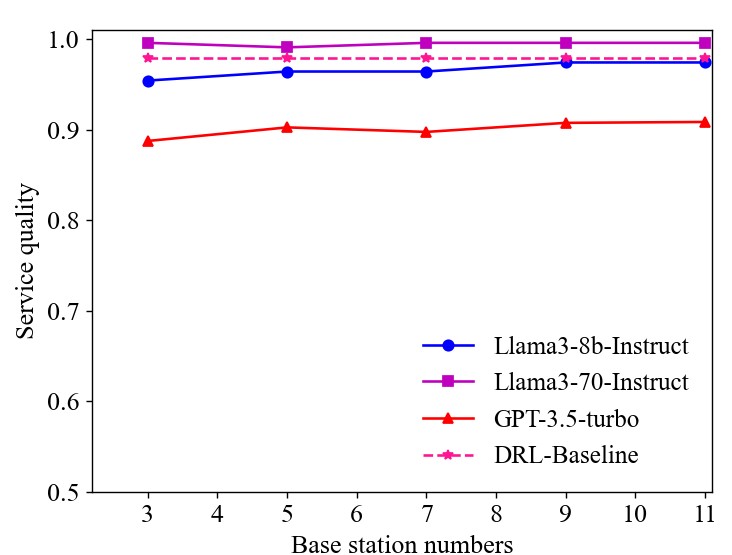} 
\caption{Continuous state space: Average service quality comparison with increasing number of BSs.}
\label{f-r7}
\end{figure}

\textbf{3) Changed Environment and Simulation Settings}:
In addition, we observe the algorithm's performance under different minimum data rate constraints. Fig. \ref{f-r4}, \ref{f-r5}, and \ref{f-r6} present the average reward, power consumption, and service quality, respectively. 
Here, every value in the following Fig. \ref{f-r4} to \ref{f-r6} is obtained by taking the average performance of converged episodes of corresponding LLMs as in Fig. \ref{f-r2} and \ref{f-r3}. 
As expected, the simulation results show that increasing the minimum data rate constraint leads to lower reward, lower service quality, and higher power consumption. 
They demonstrate that the proposed in-context learning can adapt to different optimization settings and then adjust their policies to improve the performance of target tasks.
Under various environment requirements, the proposed technique can maintain a reasonable performance compared with the existing baseline DQN.

\textbf{4) Scalability of the Proposed Algorithm}: Fig. \ref{f-r7} evaluates the scalability of the proposed algorithm by increasing network size, e.g., the number of optimized BSs. 
It shows that the proposed techniques can maintain satisfactory performance with increasing base station numbers. LLM-based techniques have the potential for large-scale network management and optimization.
The huge number of parameters of LLMs allows for large-scale information extraction, input and processing, and then generates the desired output.

\begin{figure}[h]
\centering
\includegraphics[width=8cm,height=6cm]{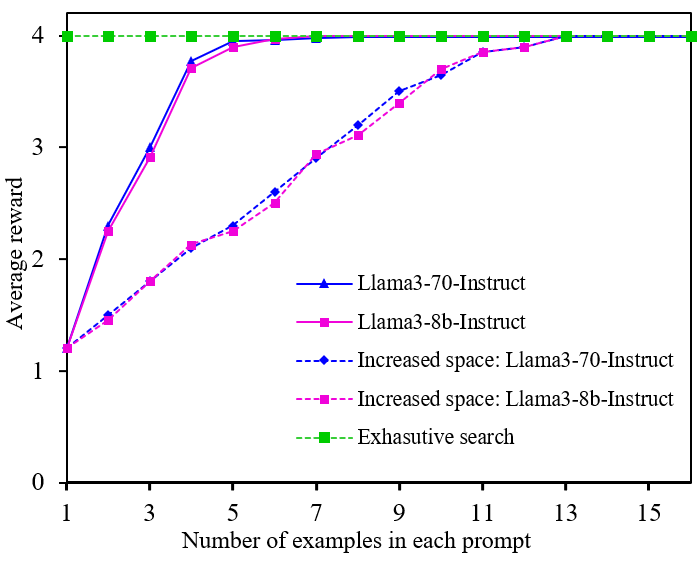}
\caption{Reward performance with increasing number of examples and larger spaces.}
\label{fig-state}
\end{figure}

\textbf{5) Effect of the number of examples}:  
In Fig. \ref{fig-state}, we evaluated the system performance with enlarged state space and changing number of examples in the prompt. Firstly, one can observe that increasing the number of examples can constantly improve the average reward. However, such improvement becomes less obvious when plenty of examples are provided.
On the other hand, increasing the state space means that more examples are needed in the prompt to achieve a satisfactory performance, e.g., more references and experience are needed to make proper decisions. 
However, it is worth noting that the overall performance is still constantly improving by increasing the number of provided examples, and it finally achieves a comparable performance as the exhaustive search method.

In summary, the above simulation results from Fig. \ref{f-r1} to Fig. \ref{f-r6} demonstrate that the proposed in-context learning technique can learn from previous explorations, optimize network performance iteratively, and adapt to different network environments. 
It achieves comparable performance as existing benchmarks DQN, and avoids the complexity of dedicated model parameter updating. 
In-context learning is considered a promising technique for future network optimization and management.
The simulation results demonstrate that the algorithm's performance is closely related to the capabilities of specific LLMs. 
For instance, Llama3 represents state-of-the-art LLM designs, while GPT-3.5 is an early LLM model.  
Therefore, it is reasonable that Llama3-8b and Llama3-70b maintain comparable performance as the DRL baseline, while GPT-3.5 turbo presents a worse performance in different tasks. 
It highlights the importance of selecting appropriate LLMs to handle various tasks.

\section{Conclusion}
LLM is a promising technique for future wireless networks, and this work proposes an LLM-enabled in-context learning algorithm for BS transmission power control.
The proposed algorithm can handle both discrete and continuous state problems, and the simulations show that it achieves comparable performance as conventional DRL algorithms. 
This work demonstrates the great potential of in-context learning for handling network management and optimization problems. 
In the future, we will consider more complicated network optimization tasks and involve more control variables and decisions.

\bibliography{References.bib}
\bibliographystyle{icml2025}

\newpage
\appendix
\onecolumn

\vspace{15pt}
\section{Input Prompt and LLM Replies Analyses}

The following presents some supplementary experiment results and discussions, including detailed input prompts and different LLMs’ replies, DRL algorithm design and implementation comparisons.  
Below is our designed prompt input to LLMs.

\begin{tcolorbox}[title = {Prompt input of LLMs for BS power control}]
\small 
You have a decision-making task for Base Station power control.

Task goal: You need to select between 4 power levels from 1 to 4. \\
\vspace{-6pt}

Task description: You have to consider the specific average user distance of each case. \\
\vspace{-6pt}

Following are some good examples I recommend: \\
\vspace{-6pt}

Example 49: Average user distance to the Base station: 8.78, your selected power level: 1.0, reward: 4.0, evaluation = good. \\ \vspace{-7pt}

Example 85: Average user distance to the Base station: 9.24, your selected power level: 1.0, reward: 4.0, evaluation=good. \\ \vspace{-8pt}

Example 58: Average user distance to the Base station: 9.19, your selected power level: 1.0, reward: 4.0, evaluation=good. \\ \vspace{-8pt}

Example 67: Average user distance to the Base station: 9.15, your selected power level: 1.0, reward: 4.0, evaluation=good. \\ \vspace{-8pt}

Example 13: Average user distance to the Base station: 9.15, your selected power level: 1, reward: 4, evaluation=good. \\ \vspace{-8pt}

Example 49: Average user distance to the Base station: 8.96, your selected power level: 1.0, reward: 4.0, evaluation=good.\\

Following are some bad examples I do not recommend: \\
\vspace{-6pt}

Example 21: Average user distance to the Base station: 10.05, your selected power level: 3, reward: 2, evaluation=bad.  \\ \vspace{-8pt}

Example 40: Average user distance to the Base station: 9.28, your selected power level: 4, reward: 1, evaluation=bad.  \\ \vspace{-8pt}

Example 120: Average user distance to the Base station: 9.9, your selected power level: 3, reward: 2, evaluation=bad.  \\ \vspace{-8pt}

Example 12: Average user distance to the Base station: 9.8, your selected power level: 3, reward: 2, evaluation=bad.  \\ \vspace{-8pt}

Example 39: Average user distance to the Base station: 10.02, your selected power level: 2, reward: 3, evaluation=bad.\\
\vspace{-8pt}

Now I will give you a new condition to solve: Average user distance to the Base station: 8.72. \\
\vspace{-6pt}

Please select from “level 1”, “level 2”, “level 3”, and “level 4” based on the above examples. Please explain the reasons for your selection.
\end{tcolorbox}

\begin{tcolorbox}[title = {Reply Sample 1 of Llama3-8b instruct.}]
\blue{Based on the provided examples}, I notice that when the average user distance is around 9.2, the recommended power level is consistently level 1. \blue{This is evident from examples 85, 58, 67, and 13}, where the average user distance is between 9.15 and 9.24, and the selected power level is always 1. \\
\vspace{-6pt}

Given the new condition: Average user distance to the Base station: 9.22, I will select 'level 1' as the recommended power level. \blue{My reasoning is that the average user distance is very close to the distances in the good examples (9.15-9.24)}, and the recommended power level in those cases was always level 1. Therefore, I believe that level 1 is the most suitable choice for this condition. \\
\vspace{-6pt}

\blue{I do not recommend selecting higher power levels (2, 3, or 4) because they were not chosen in the good examples}, and selecting them may result in a lower reward or evaluation, as seen in the bad examples.
\end{tcolorbox}

\begin{tcolorbox}[title = {Reply Sample 2 of Llama3-8b instruct.}]
\blue{Based on the provided examples}, I notice that when the average user distance to the Base station is relatively short (around 8-9 units), the \blue{recommended power level is always 1.0}, which corresponds to a \blue{"good" evaluation}.

\blue{In the new condition}, the average user distance to the Base station is 8.72, which \blue{is even shorter than the distances in the "good" examples}. Therefore, I will select power level 1.0 for this condition.

I reason that \blue{since the average user distance is relatively short, it's likely that the users are closer to the Base station and require less power to maintain a good connection}. Power level 1.0 has been consistently recommended in similar scenarios, so I believe it's the most suitable choice for this condition as well.
\end{tcolorbox}

\begin{tcolorbox}[title = {Reply Sample 3 of GPT-3.5 Turbo.}]
\blue{Based on the provided examples and conditions}:\\
\vspace{-6pt}

- \blue{Power level 1 (low power) was used in all the good examples} where the average user distance ranged from 8.78 to 9.24, resulting in a reward of 4.0 in each case.\\
\vspace{-6pt}

- Power levels 2, 3, and 4 (higher power levels) were used in the bad examples \blue{where the average user distance was greater than 9.28}, leading to lower rewards ranging from 1 to 3.\\
\vspace{-6pt}

Given that the \blue{new condition has an average user distance of 8.72}, which falls within the range of the distances \blue{in the good examples}, it would be advisable to select **level 1** for the power control. This choice is based \blue{on the pattern observed in the examples where lower power levels were more effective for shorter user distances}, resulting in higher rewards and positive evaluations.\\
\vspace{-6pt}

By selecting level 1 for this new condition, you are likely to achieve a good performance outcome similar to the ones observed in the good examples provided.
\end{tcolorbox}

The above reply samples demonstrate that LLMs have the following capabilities:
\begin{itemize}
    \item \textbf{LLMs can indeed learn from examples and then solve the target tasks}. For instance, all three reply samples start by 
    “\textit{Based on the provided examples, I notice that....}” and “\textit{Based on the provided examples and conditions}”.  
    Sample 1 also provides detailed example numbers: “\textit{This is evident from examples 85, 58, 67, and 13}”. All three models immediately frame their decision around the provided examples (“Based on the provided examples…”), showing they can ingest few-shot demonstrations and abstract a decision rule. Even Llama3-8B (Sample 1) pulls out specific indices (85, 58, 67, 13), evidencing a capacity to reference and align with particular data points.

    \item \textbf{LLMs can clearly understand the relationship between control variables, environment states, and rewards in wireless network optimization problems.} For example, in reply sample 3 of GPT-3.5 turbo, it first analyzes the selection of power level 1 and 2 , 3, 4, then explains the \textit{ “Given that the new condition has an average user distance of...”}, and finally make the decision by \textit{“it would be advisable to select **level 1** for the power control”}.  Each reply articulates how the state variable (user distance) maps to the control variable (power level) and ultimately to the reward. GPT-3.5 Turbo (Sample 3) explicitly contrasts the “good” vs. “bad” examples based on distance thresholds before recommending level 1, underscoring an understanding of the underlying optimization structure.

    \item \textbf{LLM can provide useful explanations for their decisions.} One of the key advantages of LLMs is to provide detailed explanations of their selections. For instance, sample 2 shows that “\textit{I reason that since the average user distance is relatively short, it's likely that the users are closer to the Base station and require less power to maintain a good connection}”. Such a capability is crucial when facing more complicated scenarios, providing useful references for human understanding. This is also far beyond the capabilities of existing ML techniques. Beyond selecting an action, the models generate coherent justifications. Sample 2’s justification (“users are closer…require less power”) demonstrates how LLMs can produce intuitive, domain-relevant reasoning in human-readable form—an advantage over opaque “black-box” predictors.

    \item \textbf{Structured presentation and clarity.} Finally, GPT-3.5’s bullet-point format lays out observations, contrasts, and conclusions in digestible steps, highlighting the models’ ability to organize complex decision logic in a clear, hierarchical manner—which is essential for system-level reporting and auditability.
\end{itemize}

\section{Potential Issues and Hallucination}

We also observed other replies from the Llama3-8b instruct model, and noticed some potential issues. For instance, the text box below shows the LLM replies when given another task, in which the average user distance is 8.72. 

Similar to previous replies, the LLM also clearly understands the key is to “\textit{the relationship between the average user distance and the selected power level}”. Then it analyzes the good and bad examples provided by “\textit{From the good examples}” and “\textit{From the bad examples}”. After that, it compares the target condition (average user distance is 8.72) with the good examples by “\textit{the average user distance to the Base station is 8.72, which is closer to...}”. Finally, it explains the reasons why selecting power level 1, e.g., “\textit{the average user distance is relatively close to the Base station, a lower power level (1) would be...}”. 

Although the LLM model made a good decision and obtained a satisfactory reward, it is worth noting that there is a problem when analyzing the distance. It claims that “\textit{when the average user distance is farther away, e.g., 7.76, 8.12,}”. However, 7.76 and 8.12 are obviously closer than the above distance of 8.72.

On the one hand, this experiment shows that hallucination is still one of the main problems for LLM implementations; on the other hand, in this task, we also noticed the importance of “good examples”, which have a great impact on LLM's decision-making. 
For example, the 4th paragraph demonstrates that LLM considers good examples as critical references when selecting the power level.

\begin{tcolorbox}[title = {Problematic Replies of Llama3-8b instruct.}]
Based on the provided examples, I will analyze \blue{the relationship between the average user distance and the selected power level.}  \\
\vspace{-6pt}

\blue{From the good examples}, I notice that when the average user distance is close to 8.78-8.94, the selected power level is always 1.0, which results in a high reward (4.0) and a good evaluation.  \\
\vspace{-6pt}

\blue{From the bad examples}, I see that when the average user distance is \red{farther away (e.g., 7.76, 8.12,} or 9.71), the selected power level is higher (3 or 4), which results in a lower reward (1 or 2) and a bad evaluation. \\
\vspace{-6pt}

Considering the new condition, \blue{the average user distance to the Base station is 8.72, which is closer to the range of the good examples (8.78-8.94)}. Therefore, I recommend selecting power level 1.  \\
\vspace{-6pt}

My reason is that since \blue{the average user distance is relatively close to the Base station, a lower power level (1) would be more suitable to maintain a good connection and achieve a high reward}.
\end{tcolorbox}

\end{document}